# Hypergraph Ramsey Numbers and Adiabatic Quantum Alogrithm

Ri Qu, Yan-ru Bao

*School of Computer Science & Technology, Tianjin University, Tianjin 300072, China*

Gaitan and Clark [Phys. Rev. Lett. 108, 010501 (2012)] have recently presented a quantum algorithm for the computation of the Ramsey numbers *R(m, n)* using adiabatic quantum evolution. We consider that the two-color Ramsey numbers *R(m, n; r)* for r-uniform hypergraphs can be computed by using the similar ways in [Phys. Rev. Lett. 108, 010501 (2012)]. In this paper, we show how the computation of *R(m, n; r)* can be mapped to a combinatorial optimization problem whose solution be found using adiabatic quantum evolution.

**1 Introduction**

There have already been many types of Ramsey numbers including specific graph, hypergraph, multicolor Ramsey numbers, and so on [1]. We will focus on two-color Ramsey numbers for *r*-uniform hypergrahs. *R(m, n; r)* is defined as the least integer *N* such that, in any coloring with two colors (red and blue) of *r*-subsets of the set of *N* vertices, the set of red *r*-subsets or the set of blue *r*-subsets contain an *r*-uniform complete subhypergraph. The definition can also be equivalently shown that for any integer $N$, $N \geq R(m,n;r)$ if and only if every *r*-uniform hypergraph with *N* vertices will contain an *r*-uniform complete subhypergraph with *m* vertices, or an *n*-independent set. $R(m, n; r)$ is an example of the two-color Ramsey numbers for *r*-uniform hypergrahs. In particular, $R(m,n;r) = R(m,n)$ if $r = 2$. If $\min(m,n) > r \geq 3$, the only known value of a Ramsey number *R(m, n; r)* for *r*-uniform hypergraphs is $R(4,4;3) = 13$ [2].

Ref. [3] has recently presented a quantum algorithm for the computation of the Ramsey numbers *R(m, n)* using adiabatic quantum evolution [4]. In this paper, we will show that *R(m, n; r)* for *r*-uniform hypergraphs can be computed by using the similar ways in Ref. [3].

This paper is organized as follows. In Sec. 2 we show how the computation of *R(m, n; r)* can be mapped to a combinatorial optimization problem. In Sec. 3 we use adiabatic quantum evolution to solve the combinatorial optimization problem and compute *R(m, n; r)*. We summarize the results in Sec. 4.

**2 Optimization Problem**

For given integers *N* and *r*, we can establish a 1-1 corrsepondence between the set of *r*-uniform hypergraphs with *N* vertices and $\{0,1\}^{B(N,r)}$ where *B* denotes the binomial coefficient. For convenience, let $V \equiv \{1,2,...,N\}$ be the set of *N* vertices. We define the set of *r*-uniform hypergraphs with *N* vertices by $P_N^r \equiv \wp(\{e \mid e \subseteq V \wedge |e| = r\})$ where $\wp(A)$ and $|A|$ denote the power set and the cardinal number of the set $A$ respectively. For every hypergraph $G \in P_N^r$, there exists a unique $\overbrace{N \times N \times \cdots \times N}^{r \text{ copies}}$ matrix $A(G)$ with the element

$$a_{i_1,i_2,\ldots,i_r} = \begin{cases} 1 & \{i_1,i_2,\ldots,i_r\} \in G \\ 0 & \{i_1,i_2,\ldots,i_r\} \notin G \end{cases}, \qquad (1)$$

where $i_1,i_2,\ldots,i_r \in V$. It is known that $a_{i_1,i_2,\ldots,i_r} = a_{\delta(i_1,i_2,\ldots,i_r)}$, where $\delta(i_1,i_2,\ldots,i_r)$ denotes the permutation of $i_1,i_2,\ldots,i_r$. Moreover, if $|\{i_1,i_2,\ldots,i_r\}| < r$ then $a_{i_1,i_2,\ldots,i_r} = 0$. Thus we can construct a 1-1 correspondence $g_{N,r}: P_N^r \to \{0,1\}^{B(N,r)}$ which satisfies

$$\forall G \in P_N^r, g_{N,r}(G) = a_{i_1^1,i_1^2,\ldots,i_1^r} a_{i_2^1,i_2^2,\ldots,i_2^r} \cdots a_{i_{B(N,r)}^1,i_{B(N,r)}^2,\ldots,i_{B(N,r)}^r}, \qquad (2)$$

where $\forall k, j \in \{1,2,\ldots,B(N,r)\}$, $i_k^1, i_k^2, \ldots, i_k^r \in V$, $i_k^1 > i_k^2 > \ldots > i_k^r$, and there exists $t \in \{1,2,\ldots,r\}$ such that $i_j^r = i_k^r, i_j^{r-1} = i_k^{r-1}, \ldots, i_j^{t+1} = i_k^{t+1}, i_j^t < i_k^t$ if and only if $j < k$. Thus for any $x \equiv x_1 x_2 \cdots x_{B(N,r)} \in \{0,1\}^{B(N,r)}$, $x_k = 1$ if and only if $\{i_k^1, i_k^2, \ldots, i_k^r\}$ is an edge in the hypergraph corresponds to $x$. We call $\{i_k^1, i_k^2, \ldots, i_k^r\}$ by the $k$-th edge of $x$ or the related edge of $x_k$. By (2), we can get $a_{i_1^1,i_1^2,\ldots,i_1^r} = a_{r,r-1,\ldots,1}$, $a_{i_2^1,i_2^2,\ldots,i_2^r} = a_{r+1,r-1,\ldots,1}$, ..., $a_{i_{B(N,r)}^1,i_{B(N,r)}^2,\ldots,i_{B(N,r)}^r} = a_{N,N-1,\ldots,N-r+1}$. If $r = 2$, $g_{N,2}(G) = a_{2,1} a_{3,1} \cdots a_{N,1} a_{3,2} a_{4,2} \cdots a_{N,2} \cdots a_{N,N-1}$ which is just the string shown in [3]. For $r = 3$, the binary string is

$$g_{N,3}(G) = a_{3,2,1} a_{4,2,1} \cdots a_{N,2,1} a_{4,3,1} a_{5,3,1} \cdots a_{N,3,1} \cdots a_{N,N-1,1} \\ a_{4,3,2} a_{5,3,2} \cdots a_{N,3,2} a_{5,4,2} a_{6,4,2} \cdots a_{N,4,2} \cdots a_{N,N-1,2} \cdots a_{N,N-1,N-2} \qquad (3)$$

For any $g_{N,r}(G)$, we choose $m$ vertices $S_\alpha = \{k_1, k_2, \ldots, k_m\}$ from $V$ and construct the product

$$C_\alpha = \prod_{i_1,i_2,\ldots,i_r \in S_\alpha}^{i_1 > i_2 > \ldots > i_r} a_{i_1,i_2,\ldots,i_r}. \qquad (4)$$

Note that $C_\alpha = 1$ if and only if $S_\alpha$ is corresponding to a complete $r$-uniform subhypergraph with $m$ vertices. We can get the sum

$$C_m[g_{N,r}(G)] = \sum_{\alpha=1}^{B(N,m)} C_\alpha \qquad (5)$$

which equals the number of complete $r$-uniform subhypergraphs with $m$ vertices in $G$. Subsequently, choose $n$ vertices $T_\alpha = \{k_1, k_2, \ldots, k_n\}$ from $V$ and form the product

$$I_\alpha = \prod_{v_{i_1},v_{i_2},...,v_{i_r} \in T_\alpha}^{i_1>i_2>...>i_r} \left(1-a_{i_1,i_2,...,i_r}\right). \tag{6}$$

It is clear that $I_\alpha = 1$ if and only if $T_\alpha$ is corresponding to an $n$-independent set. We can get the num

$$I_n\left[g_{N,r}(G)\right] = \sum_{\alpha=1}^{B(N,n)} I_\alpha \tag{7}$$

which equals the number of $n$-independent sets in $G$. Then we define

$$h_{m,n}\left[g_{N,r}(G)\right] \equiv C_m\left[g_{N,r}(G)\right] + I_n\left[g_{N,r}(G)\right]. \tag{8}$$

It follows that $h_{m,n}\left[g_{N,r}(G)\right] = 0$ if and only if $G$ does not contain a complete $r$-uniform subhypergraph with $m$ vertices or an $n$-independent set.

Just as Ref. [3], we can use $h_{m,n}\left[g_{N,r}(G)\right]$ as the cost function for the combinatorial optimization problem as follows. For given integers $N$, $m$, $n$ and $r$ we can find a binary string $s \in \{0,1\}^{B(N,r)}$ corresponding to the hypergraph $G_*$ with $N$ vertices that yields the global minimum of $h_{m,n}(x)$ over all $x \in \{0,1\}^{B(N,r)}$. It is clear that $h_{m,n}(s) > 0$ if and only if $N \geq R(m,n;r)$. Thus, for given integers $m$, $n$, and $r$ we begin with $N < R(m,n;r)$ which implies $\min_{x \in \{0,1\}^{B(N,r)}} h_{m,n}(x) = 0$ for $N$, and increment $N$ by 1 until we first find $\min_{x \in \{0,1\}^{B(N,r)}} h_{m,n}(x) > 0$. Then the corresponding $N$ will just equal to $R(m,n;r)$.

## 3 Quantum algorithm

The adiabatic quantum evolution (AQE) algorithm can solve the above combinatorial optimization problem [3]. According to (2), we will need $L$ qubits which respectively correspond to bits of $g_{N,r}(G)$, i.e. for any $k \in \{1,2,...,B(N,r)\}$, the $k$-th qubit is related with the $k$-th edge of $g_{N,r}(G)$. Thus we can identify $g_{N,r}(G)$ with the computational basis state $\left|g_{N,r}(G)\right\rangle$. Note that if $r \ll N$, we can get $B(N,r) = O(N^r)B(N,2)$. It follows that the AQE algorithm for $r$-uniform hypergraphs has the same quantum space complexity as one of standard graphs. As in Ref. [3], the time-dependent Hamiltonian $H(t)$ in the AQE is defined by $H(t) = (1-t/T)H_i + (t/T)H_p$, where $T$ is the evolution runtime, and adiabatic dynamics corresponds to $T \to \infty$. The Hamiltonians $H_p$ and $H_i$ are respectively defined as

$$H_p\left|g_{N,r}(G)\right\rangle = h_{m,n}\left[g_{N,r}(G)\right]\left|g_{N,r}(G)\right\rangle \text{ and } H_i = \sum_{l=1}^{L}\frac{1}{2}\left(I^l - \sigma_x^l\right).$$

*Algorithm 1*. Quantum algorithm for *R(m, n; r)*

Inputs: (i) given integers *m*, *n* and *r*. (ii) a strict lower bound *LOW* for *R(m, n; r)* which can be found in Ref. [1].

Outputs: *R(m, n; r)*.

Procedure:

(i) $LOW \rightarrow N$.

(ii) $B(N,r) \rightarrow L$.

(iii) Construct an *L*-qubit quantum system with the Hamiltonian *H(t)* and the initial state

$$\frac{1}{\sqrt{2^L}}\sum_{x=0}^{2^L-1}|x\rangle.$$

(iv) Run the AQE algorithm and measure the energy *E* at the end of execution. Note that the step must be run $k \square O(\ln[1-\delta]/\ln \varepsilon)$ times so that succeeds with probability $\delta > 1-\varepsilon$, where $1-\varepsilon$ is the probability for running the AQE algorithm once time to get right *E*.

(v) If $E = 0$ which implies $N < R(m,n;r)$, *N* is incremented by 1 and go to the step (ii). Otherwise, continue the next step.

(vi) $N \rightarrow R(m,n;r)$.

**4 Conclusions**

In this paper, we show how the computation of *R(m, n; r)* can be mapped to a combinatorial optimization problem whose solution be found using adiabatic quantum evolution. Let $\min(m,n) > r \geq 3$, then the smallest Ramsey number for *r*-uniform hypergraphs is $R(4,4;3) = 13$ which requires a *286*-qubit simulation, well beyond what can be done practically by classical computers [5, 6]. For calculation of *R(m, n; r)*, the problem Hamiltonian $H_p$ can be identified with the num expression of the products which are constructed by at most $t = \max\{B(m,r), B(n,r)\}$ *z*-Pauli operators as in Ref. [3]. It is clear that the expression is *t*-local Hamiltonian. According to the results in Ref. [3], Algorithm 1 is also in the quantum complexity class QMA [7].

**Ackownlegements**

This work was financially supported by the National Natural Science Foundation of China under Grant No. 61170178.